\documentclass[oldversion]{aa}
\usepackage{graphicx}
\usepackage{ulem}
\usepackage{threeparttable}
      \def\new#1 {{\bf #1 }}
      \def\cut#1 {\sout{#1} }

\begin{document}
\def\ffam {\hbox{$\,.\!\!^{\prime}$}}
\def\ffas {\hbox{$\,.\!\!^{\prime\prime}$}}
\def\ffM {\hbox{$\,.\!\!^{\rm M}$}}
\def\ffm {\hbox{$\,.\!\!^{\rm m}$}}
\def\ffs {\hbox{$\,.\!\!^{\rm s}$}}

\title{The inhomogeneous ISM toward PKS 1830--211 SW - A detailed view on molecular gas at a look-back 
time of 7.5\,Gyr 
           }
%       \thanks{Based on observations with the Green Bank Telescope (GBT) of the
%       National Radio Astronomy Observatory (NRAO) that is a facility of the
%       National Science Foundation (NSF) operated under cooperative agreement with
%       Associated Universities, Inc.}

%\subtitle{I. Overviewing the $\kappa$-mechanism}

\author{A.~Schulz\inst{1,2}
        \and
        C. Henkel\inst{3,4}
        \and
        K.~M. Menten\inst{3}
       \and
        S. Muller\inst{5}
       \and
        D. Muders\inst{3}
       \and
        J. Bagdonaite\inst{6}
       \and
        W. Ubachs\inst{6}}

\offprints{A. Schulz, \email{andreas.schulz@uni-koeln.de}}

\institute{Argelander-Institut f\"ur Astronomie, Universit\"at Bonn, Auf dem H\"ugel 71, 53121 Bonn, Germany
           \and 
           Institut f\"ur Physik und ihre Didaktik, Universit\"at zu K\"oln,, Gronewaldstr. 2, 50931 K\"oln, Germany
           \and
           Max-Planck-Institut f\"ur Radioastronomie, Auf dem H\"ugel 69, D-53121 Bonn, Germany
           \and
           Astron. Dept., King Abdulaziz University, P.O. Box 80203, Jeddah 21589, Saudi Arabia
           \and
           Dept. of Earth and Space Sciences, Chalmers University, Onsala Space Observatory, 43992 Onsala, Sweden
	   \and
           Dept. of Physics and Astronomy, VU University Amsterdam, De Boelelaan 1081, 1081 HV Amsterdam, Netherlands}

\titlerunning{Molecular absorption lines in PKS\,1830--211}

\authorrunning{Schulz et al.}

\date{Received: date / Accepted: (n)ever}

\abstract
{
Based on measurements with the Effelsberg 100-m telescope, a multi-line study of molecular species is 
presented toward the south-western source of the gravitational lens system PKS\,1830--211, which is by 
far the best known target to study molecular gas in absorption at intermediate redshift. Determining line 
parameters and optical depths and performing Large Velocity Gradient radiative transfer calculations, 
the aims of this study are (1) to evaluate physical parameters of the absorbing foreground gas at 
$z$ $\sim$ 0.89, in particular its homogeneity, and (2) to monitor the spectroscopic time variability  
caused by fluctuations of the $z$ $\sim$ 2.5 background continuum source. We find, that the gas is quite 
inhomogeneous with $n$(H$_2$) $\sim$ 2 $\times$ 10$^3$\,cm$^{-3}$ for most molecular species but with 
higher values for H$_2$CO and lower ones for SO. Measuring the CS $J$=1$\leftarrow$0 transition during a 
time interval of more than a decade, from 2001 to 2012, the peak absorption depth of the line remains 
approximately constant, while the line shape undergoes notable variations. Covering the time between 
1996 and 2013, CS, HCO$^+$, and CH$_3$OH data indicate maximal integrated optical depths in $\sim$2001 and 
2011/2012. This is compatible with a $\sim$10\,yr periodicity, which, however, needs confirmation by substantially 
longer time monitoring. Comparing molecular abundances with those of different types of Galactic and nearby 
extragalactic clouds we find that the observed cloud complex does not correspond to one particular type but 
to a variety of cloud types with more diffuse and denser components as can be expected for an observed region 
with a transverse linear scale of several parsec and a likely larger depth along the line-of-sight. 
A tentative detection of Galactic absorption in the c-C$_3$H$_2$ 1$_{10}-1_{01}$ line at 18.343\,GHz is also 
reported.}


\maketitle

%________________________________________________________________

\section{Introduction}

To study molecular cloud properties in far distant galaxies, low spatial resolution becomes an increasingly 
important problem since every observed position is averaging over large areas. Furthermore beam filling 
factors are small causing low flux densities and, therefore, low signal-to-noise ratios. A suitable 
way to overcome these problems is to use molecular lines in absorption in front of compact background continuum 
sources such as radio quasars whose small angular sizes determine the effective beam size of the observation 
and whose continuum flux densities determine the sensitivity of the measurements. The availability of such 
background-foreground pairs and the required strength of the background continuum limit the application of this 
method to a very small number of objects. The so far brightest and most distant such pair is PKS\,1830--211 
(Wiklind \& Combes 1996a), a gravitational lens with truly unique properties. The background source is likely a 
blazar at $z$ = 2.507 (Lidman et al. 1999, De Rosa et al.  2005). The foreground lens is a spiral galaxy 
observed almost face-on at redshift $z$ $\sim$ 0.89 (Courbin et al. 2002; Winn et al. 2002). It splits the 
continuum radiation of the blazar into three main image components. These consist of a stronger north-eastern 
``hotspot'', a weaker south-western ``hotspot'', and an Einstein ring being prominent only at low frequencies 
($<$10\,GHz; Jauncey et al. 1991). The SW source is located behind a spiral arm of the lensing galaxy, at a 
distance of about 2 kpc from its center (Muller et al. 2006), and shows by far the strongest molecular 
absorption lines among all known sources of its class (for the others, see Wiklind \& Combes 1994, 1995, 1996b 
and Kanekar et al. 2005). 

The exceptional situation concerning continuum flux and line strength encountered in PKS\,1830--211 has been 
extensively explored, leading to a large number of detected molecular lines and species (e.g., Henkel 
et al. 2009; Muller et al. 2011, 2014a,b). Therefore it provides detailed insight into the properties of the 
interstellar medium of a galaxy at an intermediate cosmological distance. Adopting a standard $\Lambda$-cosmology 
with H$_0$ = 67\,km\,s$^{-1}$\,Mpc$^{-1}$, $\Omega_{\rm m}$ = 0.315, and $\Omega_{\Lambda}$ = 0.685 (Planck 
Collaboration et al. 2014), $z$ = 0.88582 corresponds to a lookback time of about 7.5\,Gyr, more than half 
the present age of the universe. 

It is the aim of this article to examine physical properties of the molecular gas using molecular lines 
characterizing various physical conditions. This includes a first evaluation of the homogeneity of the 
absorbing molecular gas column and CS $J$ = 1$\leftarrow$0 spectra covering more than a decade to study
variability caused by the background continuum source.

\section{Observations}

The observations were carried out in a position switching mode with the primary focus $\lambda$ = 1.9, 1.3, 
and 1.0\,cm receivers of the 100-m telescope at Effelsberg/Germany\footnote{The 100-m telescope at Effelsberg is operated 
by the Max-Planck-Institut f{\"u}r Radioastronomie (MPIfR) on behalf of the Max-Planck-Gesellschaft (MPG).}, mostly 
within the short period between August and December 2001. On- and off-source integration times were two minutes per
scan, yielding full cycles of 4--5 minutes duration. For full width to half power (FWHP) beam widths, system temperatures, 
and aperture efficiencies, see Table~\ref{tab1}. During 2001--2006, the backend was an 8192 channel autocorrelator, 
which was split into eight segments, covering selected frequency ranges within a total bandwidth of 500\,MHz. Each of 
the eight segments had a bandwidth of 40\,MHz with 512 channels, yielding channel spacings of 0.7--1.7\,km\,s$^{-1}$. 
At $\lambda$ $\sim$ 1.3\,cm, but not at 1.9 and 1.0\,cm, a dual channel receiver was used sampling orthogonal linear 
polarizations. Therefore, in this case at least two of the eight segments were centered at the same frequency to maximize 
sensitivity by observing simultaneously the two polarization components. In 2006, a Fast Fourier Transform Spectrometer 
(FFTS) was employed with a bandwidth of 100\,MHz and 16384 channels for each polarization channel. For the more recent 
CS data from April 2011 to April 2012 another FFTS was used, this time with a bandwidth of 2\,GHz and 32768 
channels per polarization. At 26\,GHz, the corresponding FFTS channel spacings are 0.07 and 0.7\,km\,s$^{-1}$, 
respectively. For recent (2012/3) methanol measurements, the 32768 channel FFTS has also been used, but in this case
with a bandwidth of 100\,MHz. Details of these recent methanol observations can be found in Bagdonaite et al. (2013a,b).

As already mentioned (Sect.\,1), PKS\,1830--211 is with $\sim$5\,Jy near $\lambda$ = 1\,cm one of the strongest 
continuum sources in the radio sky. Therefore, focus and pointing could be checked toward the source itself. The latter 
was found to be accurate to about 5\arcsec. The continuum was also used to calibrate the spectral lines to obtain 
line-to-continuum flux density ratios, which could be determined with an accuracy of up to $\pm$7\% depending 
on the strength of the respective line (see also Sect.\,4). Data reduction was obtained using standard procedures 
of CLASS and GREG, two tools of the GILDAS\footnote{Grenoble Image and Line Data Analysis Software, see 
http://www.iram.fr/IRAMFR/GILDAS} software package. During the data reduction, only linear baselines were subtracted 
from the original spectra.

\begin{table}
\begin{threeparttable}
\caption[]{Observational parameters.}
\label{tab1}
\begin{flushleft}
\begin{tabular}{ccccc}
\hline 
     $\nu$ & $\lambda$ & $\theta_{\rm b}^{\rm a)}$ & $T_{\rm sys}^{\rm b)}$ & $\eta_{\rm A}^{\rm c)}$ \\
     (GHz) &   (cm)    &           ($''$)          &     (K)                &                         \\
\hline
           &           &                           &                        &                         \\
13.5--18.7 &   1.9     &  60--45                   &      40                &   0.39--0.32            \\
18.0--26.0 &   1.3     &  45--35                   &      70                &   0.33--0.26            \\
27.0--36.7 &   1.0     &  32--23                   &      75                &   $\sim$0.30            \\
           &           &                           &                        &                         \\
\hline
\end{tabular}
\begin{tablenotes}
\item[a)] Full Width to Half Power (FWHP) beam width in arcsec. 
\item[b)] System temperatures in units of antenna temperature ($T_{\rm A}^*$). 
\item[c)] Aperture efficiency.
\end{tablenotes}
\end{flushleft}
\end{threeparttable}
\end{table}

\begin{figure}[t]
\vspace{-0.0cm}
\hspace{1.0cm}
%\centering
%\resizebox{19.0cm}{!}{\rotatebox[origin=br]{-90}{\includegraphics{as-figure1.ps}}}
\resizebox{7cm}{!}{\includegraphics[trim=30mm -10mm 10mm 0mm]{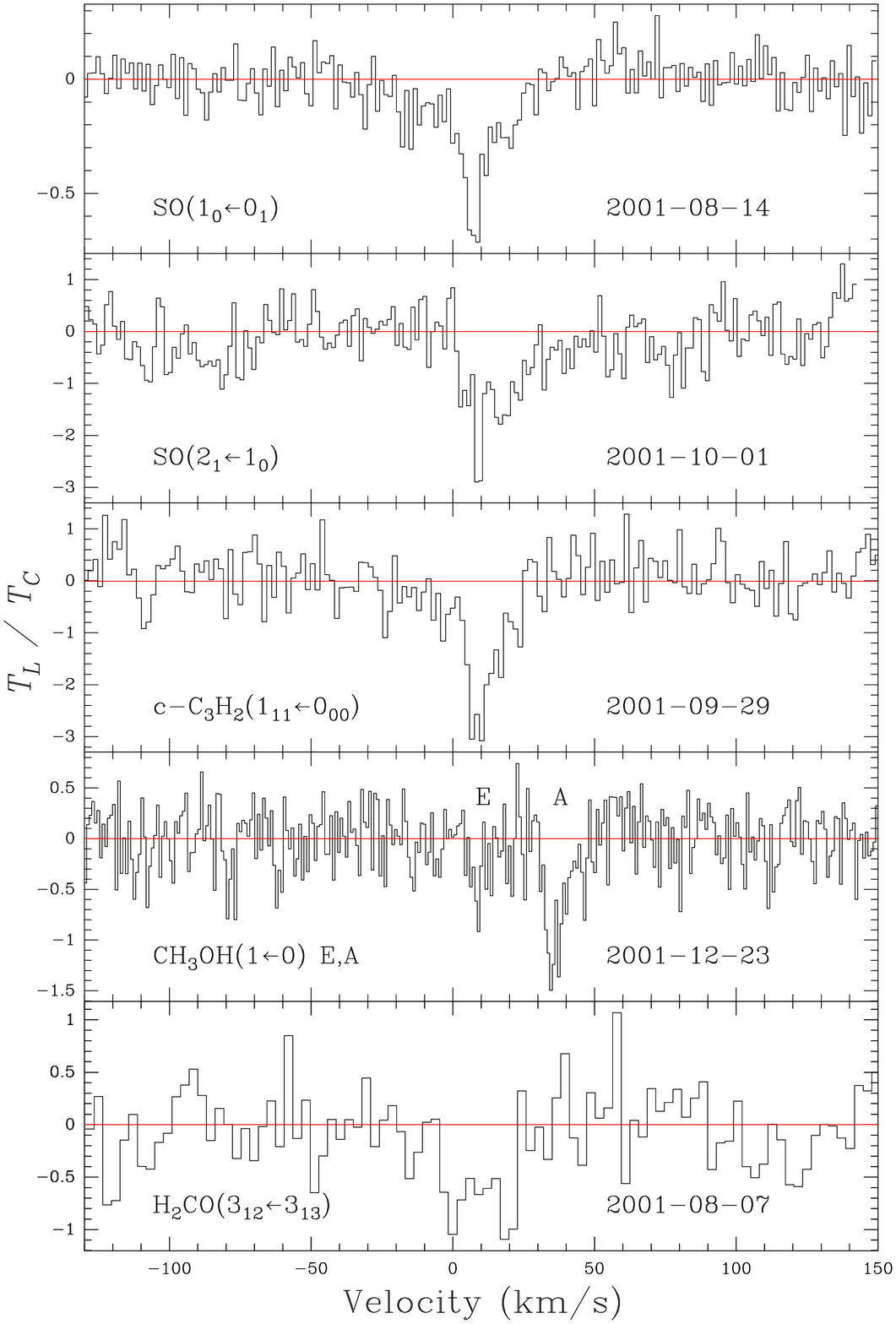}}
\vspace{-0.7cm}
\caption{The six absorption lines detected toward PKS\,1830--211, displayed in five panels with a Local Standard of 
Rest (LSR) velocity scale relative to $z$=0.88582 ($V_{\rm LSR}$ = $V_{\rm HEL}$ + 12.4\,km\,s$^{-1}$; 
ordinate: absorption depth in units of percent of the observed continuum flux density, $T_C$, accounting
for the entire source). The spectra were taken between August and December 2001. The velocity scale of the fourth 
panel from the top is based on the E-transition of methanol (CH$_3$OH). The scale for the A-transition
is offset by --27.5\,km\,s$^{-1}$. The cyclic-C$_3$H$_2$ line of the central panel belongs to the 
para species (see Table~2). Channel spacings are 1.4, 1.4, 1.7, 0.9 and 3.0\,km\,s$^{-1}$ from top to bottom.} 
\label{fig1}
\end{figure}

\section{Results}

\begin{figure}[t]
\vspace{-0.0cm}
\centering
%\resizebox{19.0cm}{!}{\rotatebox[origin=br]{-90}{\includegraphics{as-figure2.ps}}}
\resizebox{8.5cm}{!}{\rotatebox[origin=br]{-90}{\includegraphics[trim=0mm 0mm -25mm 0mm]{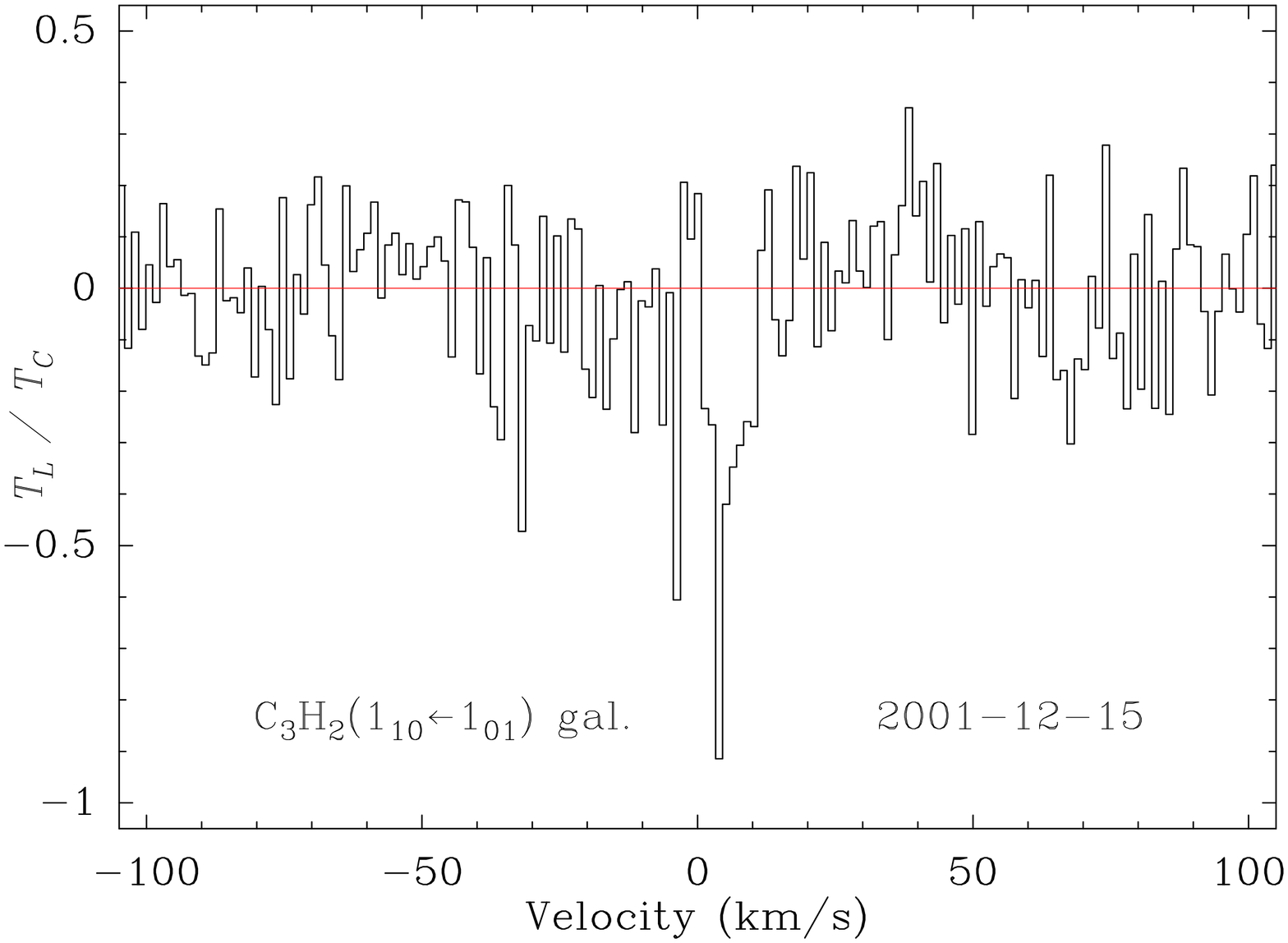}}}
\vspace{-0.7cm}
\caption{The tentative detection of Galactic c-C$_3$H$_2$ (cyclic variant, ortho-species), 
with a channel spacing of 1.3\,km\,s$^{-1}$. The spectrum was taken on December 15, 2001. The ordinate 
denotes absorption in units of percent of the entire observed continuum flux density, $T_C$, of 
PKS\,1830--211. The abscissa is Local Standard of Rest velocity for redshift zero.}
\label{fig2}
\end{figure}

The separation between the north-eastern and south-western hotspots of PKS\,1830--211 (see Sect.\,1) is about 1\,arcsec. 
Therefore, they are clearly within our much larger beam sizes (Table 1). Figure~\ref{fig1} shows the line profiles 
measured at Effelsberg between August and December 2001, with the ordinate displaying absorption in units of percent 
of the total observed continuum flux density. The abscissa depicts Local Standard of Rest (LSR) velocity relative to $z$ = 
0.88582. The detected absorption lines originate all from the SW hotspot. We have obtained five spectra (one containing 
two lines) of the molecules SO, para-c-C$_3$H$_2$, CH$_3$OH, and ortho-H$_2$CO. While, in the meantime, these 
molecules have also been detected in other studies (e.g., Muller et al. 2011), these are the earliest detections except 
for H$_2$CO and C$_3$H$_2$ (Menten et al. 1999). The spectra do not show absorption line components at an LSR velocity 
of about {\hbox{--135\,km\,s$^{-1}$}}, which might appear in front of the NE hotspot (Wiklind \& Combes 1998, Muller et al. 
2006). Fig.~\ref{fig2} depicts our tentative detection of Galactic C$_3$H$_2$, which would be the first detection of Galactic 
molecular gas absorption toward the line-of-sight to PKS\,1830--211. 

Apparent optical depths are determined using  
$$
\tau_{\rm app} = -{\rm ln}\left(1 - \frac{|T_{\rm L}|}{T_{\rm C}} \right) 
$$
(e.g., Henkel et al. 2009), with $T_{\rm L}$ and $T_{\rm C}$ denoting line and continuum temperature and assuming 
(realistically) that the excitation temperature $T_{\rm ex}$ is negligible relative to $T_{\rm C}$ (see also Sect.\,4.4).  
Line parameters and apparent optical depths are given in Table~\ref{tab2}. The table also contains upper limits to
the optical depths of six molecular transitions and two recombination lines, also obtained between August and December 
2001, and additional data taken from the literature. The noise level of the data with detected absorption does not allow 
us to discriminate between several velocity components as it has been done, for example, by Henkel et al. (2009) when 
analyzing their CS\,$J$=1$\leftarrow$0 profile.

\begin{table*}
\caption[]{Line parameters and derived optical depths}
\label{tab2}
\label{linepar}
\begin{tabular}{lccrrcccrl}
\hline
Transition & $\nu_{observed}$ & $\nu_{rest}$ & $V$  & ${\Delta V}$ &  $f_{\rm c}$  &  ${\tau}$  &  ${\tau}$  &  ${\Delta \tau}$  &  Remarks  \\
              & \multicolumn{2}{c}{(GHz)} &  \multicolumn{2}{c}{(km\,s$^{-1}$)} &        &            &  $f_{\rm c}$= .38      &             &          \\
\hline
                                      &             &          &              &               &        &           &            &            &          \\
				      &             &          & \multicolumn{3}{c}{\bf Our Detections}&           &            &            &          \\
$\rm H_2CO\ 3_{12}\leftarrow3_{13}$(o)& 15.36456    & 28.97480 &  8.9$\pm$2.7 &  27.1$\pm$5.3 &  .153  & .054      & .021      & 30\%        & 070801   \\
$\rm SO\ 1_0\leftarrow0_1$            & 15.90902    & 30.00154 &  7.5$\pm$0.4 &   5.4$\pm$1.2 &  .156  & .031      & .013      & 10\%        & 140801, 1\\
				      &             &          &  5.8$\pm$2.4 &  37.2$\pm$5.8 &  .156  & .016      & .006      & 25\%        & 140801, 1\\
$\rm c-C_3H_2\ 1_{10}\leftarrow1_{01}$(o) & 18.34314& 18.34314 &  4.6$\pm$0.6 &   5.1$\pm$1.5 &  1.00  & .007      & ---       & $\ga$ 30\%  & 151201, 2\\
$\rm CH_3OH\ 1_{0}\leftarrow0_{0} (A^+)$& 25.65063  & 48.37247 &  8.7$\pm$0.5 &   8.3$\pm$1.4 &  .196  & .065      & .033      & 20\%        & 231201, 3\\
$\rm CH_3OH\ 1_{0}\leftarrow0_{0} (E)$& 25.65297    & 48.37689 &  8.7$\pm$1.3 &   5.1$\pm$4.9 &  .196  & .020      & .010      &100\%        & 231201, 3\\
$\rm c-C_3H_2\ 1_{11}\leftarrow0_{00}(p)$& 27.49012 & 51.84141 & 10.2$\pm$0.7 &  15.4$\pm$2.3 &  .198  & .097      & .050      & 20\%        & 290901   \\
$\rm SO\ 2_1\leftarrow1_0$            & 33.37105    & 62.93179 &  8.6$\pm$1.1 &  20.9$\pm$2.5 &  .227  & .063      & .038      & 20\%        & 011001   \\ 
\hline
                                      &             &          &              &               &        &           &            &            &          \\
				      &             &          & \multicolumn{3}{c}{\bf Our Nondetections} &       &            &            &          \\
$\rm HC_5N\ 13\leftarrow12$           & 18.35508    & 34.61439 &      --      &     --        & .17    & $<$.003   & $<$.001    &  7\%       & 151201, 4\\
$\rm SiS\ 2\leftarrow1$               & 19.25403    & 36.30963 &      --      &     --        & .17    & $<$.002   & $<$.001    &  7\%       & 151201, 4\\
H56$\alpha$                           & 19.33709    & 36.46627 &      --      &     --        & .17    & $<$.018   & $<$.008    &  7\%       & 151201, 4\\
H70$\beta$                            & 19.49376    & 36.76172 &      --      &     --        & .17    & $<$.018   & $<$.008    &  7\%       & 151201, 4\\
$\rm HNCO\ 2_{02}\leftarrow1_{01}$    & 23.31249    & 43.96300 &      --      &     --        & .19    & $<$.004   & $<$.002    &  7\%       & 301201, 4\\
$\rm HC_5N\ 20\leftarrow19$           & 28.23830    & 53.25235 &      --      &     --        & .21    & $<$.005   & $<$.003    &  7\%       & 031001, 4\\
$\rm SO_2\ 2_{11} \leftarrow 2_{02}$  & 28.38493    & 53.52887 &      --      &     --        & .22    & $<$.006   & $<$.004    &  7\%       & 031001, 4\\
$\rm SO\ 3_4 \leftarrow 3_3$          & 35.01657    & 66.03494 &      --      &     --        & .23    & $<$.009   & $<$.006    &  7\%       & 021001, 4\\
				      &             &          &              &               &        &           &            &            &          \\
\hline
\hline
\hline
                                      &             &          &              &               &        &           &            &            &          \\
		     		      &         & & \multicolumn{3}{c}{{\bf Menten et al. (1999)}}     &           &            &            &          \\
$\rm H_2CO\ 2_{11} \leftarrow 2_{12}$(o)& 07.68285  & 14.48848 &    $\sim$9   &  $\sim$20     & .20    & .050      & .026       & 15\%       & 5        \\
$\rm c-C_3H_2\ 2_{12} \leftarrow 1_{01}$ (o) &45.25294&85.33891&    $\sim$9   &  $\sim$20     & .35    & .210      & .190       & 15\%       & 5, 7     \\
                 		      &             &          &              &               &        &           &            &            &          \\
\hline
                                      &             &          &              &               &        &           &            &            &          \\
				      &         & &  \multicolumn{3}{c}{{\bf Muller et al. (2011)}}    &           &            &            &          \\
$\rm c-C_3H_2\ 3_{31}\leftarrow3_{22}$(p) & 31.58196& 59.55789 &  --          &  --           & .22    & .005      & .003       & 20\%       & 6        \\
$\rm CH_3OH\ 1_{0}\leftarrow2_{-1}$ (E)& 32.09821   & 60.53146 & 7.1$\pm$0.5  & 21.2$\pm$1.2  & .22    & .033      & .019       & 10\%       & 6        \\
$\rm SO\ 2_1\leftarrow1_0$            & 33.37105    & 62.93180 &11.4$\pm$0.3  & 19.7$\pm$0.7  & .23    & .050      & .030       & 10\%       & 6        \\
$\rm c-C_3H_2\ 2_{02}\leftarrow1_{11}$(p) & 43.53202& 82.09356 &  --          &  --           & .26    & .084      & .057       & 10\%       & 6        \\
$\rm c-C_3H_2\ 3_{12}\leftarrow3_{03}$(o) & 43.99476& 82.96621 &11.5$\pm$0.1  & 18.7$\pm$0.2  & .26    & .031      & .022       & 10\%       & 6        \\
$\rm c-C_3H_2\ 2_{12}\leftarrow1_{01}$(o) & 45.25294& 85.33891 &  --          &  --           & .26    & .285      & .189       & 10\%       & 6, 7     \\
$\rm SO\ 2_2\leftarrow1_1$            & 45.65332    & 86.09395 &  --          &  --           & .27    & .004      & .003       & 35\%       & 6        \\
				      &             &          &              &               &        &           &            &            &          \\
\hline

\end{tabular}
\smallskip \\
``Our Detections'' and ``Our Nondetections'' were all obtained between August and December 2001. \\
Col.\,(1): Molecular line, (o)=ortho, (p)=para; Col.\,(2): redshifted frequency, adopting $z$ = 0.88582; 
Col.\,3: Approximate rest frequency; Cols.\,(4) and (5): Local Standard of Rest velocities and full width to 
half power (FWHP) line widths, in case of SO 1$_0$$\leftarrow$0$_1$ for both the broad and the narrow component; 
Cols.\,(6) and (7): continuum source covering factor (see also Sect.\,4.1 and Henkel et al. 2009) and corresponding 
peak optical depth(s); Col.\,(8): peak optical depths for a continuum source covering factor of 0.38; Col.\,(9): 
estimated 1$\sigma$ uncertainties in percent of the optical depth. \\
Remarks (last column): \\
The 6-digit numbers provide the epoch of our observations with day, month, and year from left to right. \\
1: Possibly two line components. \\
2: Galactic line, tentative detection. A continuum source covering factor of unity has been adopted. \\
3: E- and A-transitions seen in the same spectrum (Fig.\ref{fig1}). The intensities are consistent with a line strength ratio of 1:3.8. \\
4: Upper 1$\sigma$ limits of $\tau$; spectra were smoothed to $\sim$10 km\,s$^{-1}$ channels. \\
5: Extrapolated continuum flux densities: 7.1\,Jy (14 GHz) and 3.5\,Jy (45 GHz). \\
6: Calculated from Table~10 of Muller et al. (2011). \\ 
7: $\tau$ values of Menten et al. (1999) and Muller et al. (2011) agree well. \\
%%(1): $f_{\rm c}$ after Henkel et al. (2009), see also text; (2): two line components, see text; (3): galactic line; (4) E- and A-component of 
%%this line; (5): upper limits of $\tau$, spectra smoothed to $\sim$10 km\,s$^{-1}$; (6): extrapolated continuum flux 7.1 Jy (14 GHz) and 3.5 
%%Jy (45 GHz); (7): calculated from values of their Table 10, $f_{\rm c}$ = .38; (8) $\tau$ values of Menten et al. and Muller et al. are 
%%identical. 
\end{table*}

Within the limits of noise, all the main line components obtained in 2001 show similar velocities (see also 
Henkel et al. 2008 and 2009 for additional lines at $\lambda$ $\sim$ 1--2\,cm). The Local Standard of Rest (LSR) velocities of 
the peaks of the absorption (LSR velocities will be used throughout the paper), are close to $\sim$9--10\,km\,s$^{-1}$. 
Fig.~\ref{fig3} displays CS $J$ = 1$\leftarrow$0 spectra from the south-western component taken over a time interval 
of more than a decade, from 2001 to 2012. Table~\ref{tab3} summarizes the corresponding line parameters.

\begin{table}
\begin{threeparttable}
\caption[]{CS $J$ = 1$\leftarrow$0 one component Gaussian fit parameters$^{\rm a})$} 
\label{tab3}
\begin{tabular}{cccc}
\hline 
Epoch         & $\int{(T_{\rm L}/T_{\rm c}) {\rm d}V}$ & $V_{\rm LSR}$  & $\Delta V$        \\
   	      & \multicolumn{3}{c}{(km\,s$^{-1}$)}   \\
\hline
              &                                        &                &                    \\
Dec. 15, 2001 & --4.43$\pm$0.07                        & 10.6$\pm$0.2   & 23.5$\pm$0.5       \\
Dec. 22, 2001 & --4.42$\pm$0.11                        & 09.8$\pm$0.3   & 24.9$\pm$0.8       \\
Dec. 30, 2001 & --4.38$\pm$0.05                        & 11.3$\pm$0.1   & 22.1$\pm$0.3       \\
Mar. 11, 2002 & --4.76$\pm$0.05                        & 09.6$\pm$0.1   & 26.4$\pm$0.3       \\
Sep. 27, 2002 & --4.42$\pm$0.06                        & 11.0$\pm$0.1   & 23.4$\pm$0.4       \\
Jan. 05, 2003 & --3.57$\pm$0.04                        & 11.5$\pm$0.1   & 19.1$\pm$0.2       \\
Mar. 18, 2006 & --3.13$\pm$0.04                        & 10.9$\pm$0.1   & 15.5$\pm$0.2       \\
Apr. 01. 2011 & --3.32$\pm$0.04                        & 09.9$\pm$0.1   & 18.3$\pm$0.3       \\
Dec. 08, 2011 & --4.28$\pm$0.03                        & 10.3$\pm$0.1   & 19.4$\pm$0.1       \\
Apr. 08, 2012 & --4.60$\pm$0.02                        & 10.1$\pm$0.1   & 20.0$\pm$0.1       \\
              &                                        &                &                    \\
\hline
\end{tabular}
\begin{tablenotes}
\item[a)]  Adopted redshifted frequency: $\nu$ = 25.978596\,GHz. Note that the absorption peak is 
slightly (typically 1.0--2.5\,km\,s$^{-1}$) blue-shifted with respect to the velocities given in Col.\,3, caused 
by the asymmetry in the overall lines (Fig.~\ref{fig3}). For integrated optical depths, see Fig.~\ref{fig4}.
\end{tablenotes}
\end{threeparttable}
\end{table}

\section{Analysis}

\subsection{Source covering factor and time variability}

In order to determine the true optical depths of our spectra from the south-western source, an important problem 
to be discussed is the background source coverage factor $f_{\rm c}$. Muller et al. (2006) find their HCO$^+$ feature 
at 94 GHz being saturated at a depth of 38\% relative to the total continuum level implying $f_{\rm c}$ = 0.38 at this 
frequency (see also Mart\'{\i}-Vidal et al. 2013 and Muller et al. 2014b). Trying to account for a larger extent 
of the background continuum at lower frequencies, Henkel et al. (2009) proposed a frequency dependence of this factor, 
$f_{\rm c}$ = 0.2 $\times$($\nu_{\rm observed}$/26\,GHz)$^{0.5}$ for $\nu_{\rm observed}$ $\la$ 100\,GHz, matching 
that determined by Muller et al. (2006) at the high frequency end. Therefore, we present calculated $\tau_{\rm peak}$ 
values using both, the constant coverage factor of 0.38 and the frequency-dependent factor in our Table~\ref{tab2}. 
Nevertheless, this affects our analysis only marginally since below we will use ratios of optical depths which 
lead to results that are quite independent of coupling values because frequencies are not too discrepant. 

Another problem discussed in the literature are concerns about the variability of the source (e.g., Muller \& Gu{\'e}lin 2008). 
Henkel et al. (2009) find that the depth of the absorption features is well correlated with the continuum flux. Since all our 
observations presented here for the first time with the notable exception of CS $J$ = 1$\leftarrow$0 were carried out 
during a short period within 2001 (see also Sect. 2 and Table~2), being confined to epochs within the observing period 
covered by Henkel et al. (2009), variations in the ratio of line intensity to continuum flux remain negligible.

A further important point is whether this argumentation also holds when using data taken over a longer time span. This seems
to be the case because two lines relevant for this study were observed at two largely different epochs: (1) The ortho c-C$_3$H$_2$ 
2$_{12}$$\leftarrow$1$_{01}$ transition, reported by Muller et al. (2011), was also observed by Menten et al. (1999). The agreement 
in peak optical depths is excellent (see Table~\ref{tab2}). (2) The SO 2$_1$$\leftarrow$1$_0$ line observed by us in 2001 was also
observed by Muller et al. (2011), yielding again, within the error limits, consistent results. We have to note, however, 
that this agreement may be fortuitous and would not necessarily hold, if we would compare data from other epochs.

\begin{figure}[t]
\vspace{-0.0cm}
\centering
\resizebox{8.5cm}{!}{\includegraphics[trim=0mm -15mm 0mm 0mm]{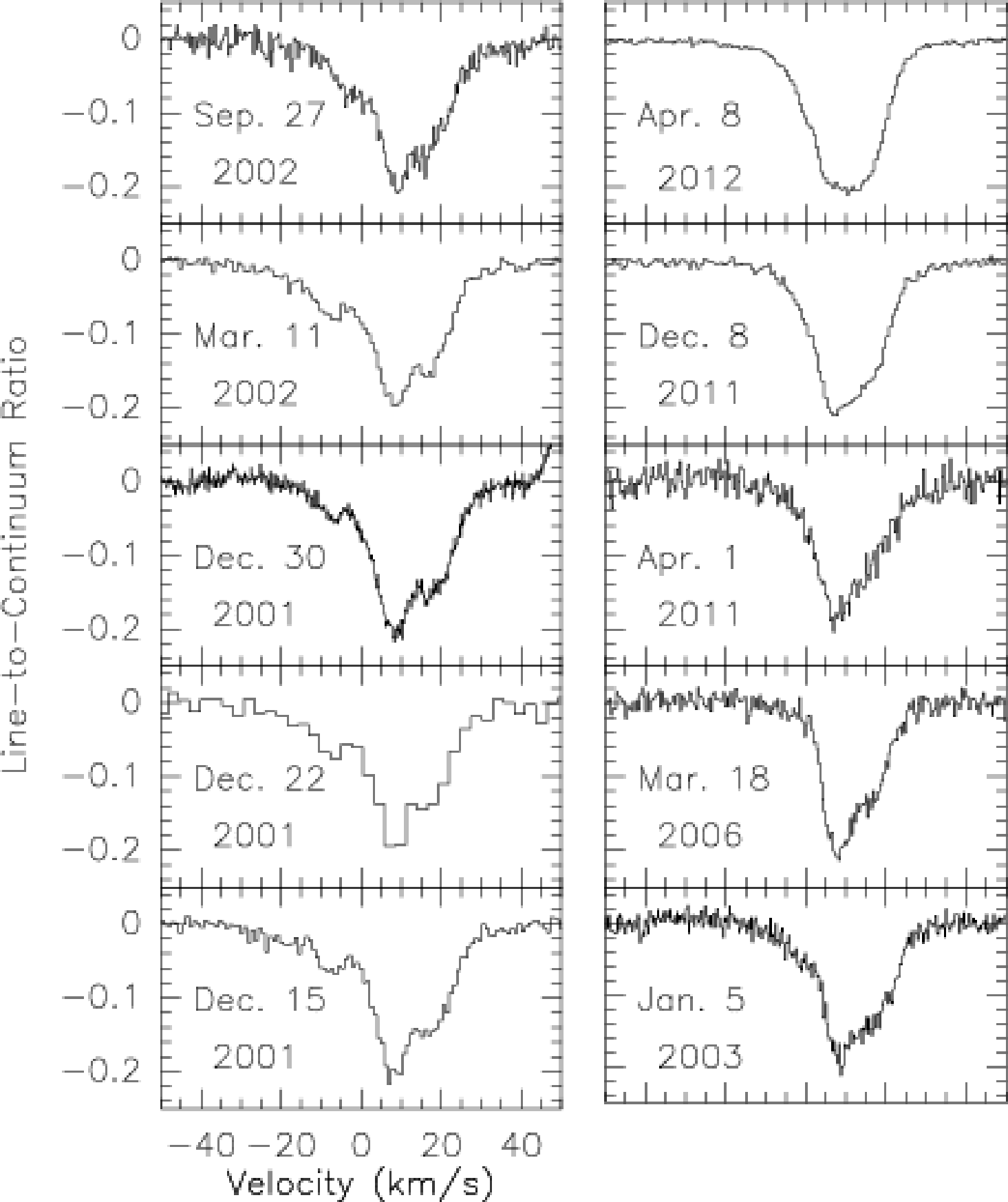}}
\vspace{-0.7cm}
\caption{Monitoring observations of the CS\,(1$\leftarrow$0) line near 26\,GHz, covering more than a decade (from
lower left to upper right). Channel spacings are (from the earliest to the latest epoch) 0.90, 2.7, 0.22, 0.90, 
0.45, 0.34, 0.45, 0.42, 0.70, and 0.70\,km\,s$^{-1}$. The ordinate denotes absorption in units of the entire 
continuum flux density of PKS\,1830--211. The abscissa is Local Standard of Rest velocity for redshift $z$ = 0.88582.}
\label{fig3}
\end{figure}

To better illustrate this effect, Fig.~\ref{fig3} shows a series of CS\,(1 $\leftarrow$ 0) lines measured between Dec.~2001 
and Apr.~2012. The depth of the absorption remains constant to within the calibration uncertainties (see Sect.\,2), while the 
line shape is slowly varying. The component at about --10\,km\,s$^{-1}$ fades in 2002. The one at about +18\,km\,s$^{-1}$ 
disappears between 2006 and 2011. In spring 2012, the profile becomes almost Gaussian, suggesting that the velocity range with 
a significant optical depth (larger than unity) has increased. Comparing these data with the observations by Muller et al. (2008) 
of the even more saturated HCO$^+$ $J$ = 2$\leftarrow$1 line (their data are presented on a heliocentric velocity scale) also 
shows a roughly constant absorption depth (implying that $f_{\rm c}$ remains approximately constant in time, see also 
Muller et al. 2006, Henkel et al. 2009), the fading of the --10\,km\,s$^{-1}$ component, and a trend to a more Gaussian profile 
near the end of their monitoring period (May~2007).

\begin{figure}[t]
\vspace{-0.0cm}
\centering
\resizebox{8.5cm}{!}{\includegraphics[trim=0mm -15mm 0mm 0mm]{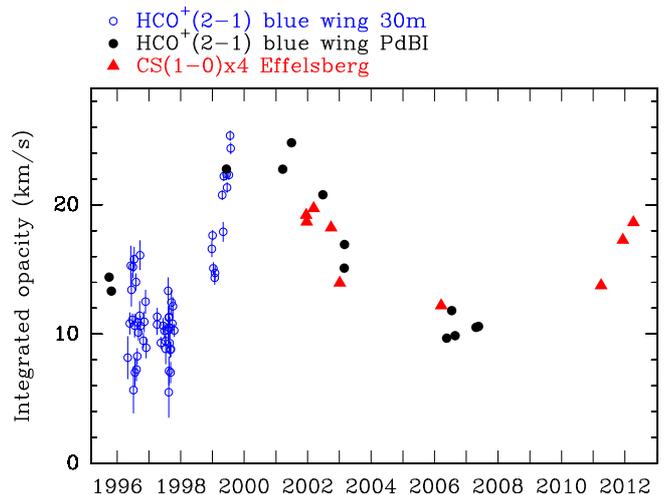}}
\vspace{-0.7cm}
\caption{Integrated opacities for $f_{\rm c}$ = 0.38 of the blue wing of the HCO$^+$ $J$ = 2$\leftarrow$1 lines
(--37.6\,km\,s$^{-1}$ $<$ $V_{\rm LSR}$ $<$ --2.6\,km\,s$^{-1}$; $V_{\rm LSR}$ = $V_{\rm HEL}$ + 12.4\,km\,s$^{-1}$;
$z$ = 0.88582). The HCO$^+$ spectra were taken from IRAM Pico Veleta (30-m) and Plateau de Bure (PdBI) data
published by Muller \& Gu{\'e}lin (2008). Their calibration uncertainties are 15\% and 1\%, respectively. For the 
Effelsberg data, see Sect.\,2.}
\label{fig4}
\end{figure}

To visualize the changes, Fig.~\ref{fig4} displays the integrated optical depths of two main molecular tracers, 
HCO$^+$ $J$ = 2$\leftarrow$1 and CS 1$\leftarrow$0, the latter amplified by a factor of four, to match the other data. Since 
HCO$^+$ shows higher optical depths than CS (see Muller \& Gu{\'e}lin 2008; Henkel et al. 2009; Muller et al. 2014b), the 
blue wings of the HCO$^+$ profiles have been plotted, which show significant variablity, indicating like CS only moderate saturation. 
The observed trend in integrated line opacity variations shows excellent agreement. To expand the monitored time interval 
further, we have also included recently taken methanol (CH$_3$OH) data from Effelsberg (see also Sect.\,2), which have been 
published by Bagdonaite et al. (2013a,b) and which show a significant decrease in integrated opacities from a peak value reached 
at the end of 2011 or early 2012. Combining these data, we see peaks in 2000--2002 and 2011--2012 and lower values at other 
times, when considering the time interval between 1995 and 2013. This suggests a period of variability of order 10\,yr. Of 
course, having obtained data from less than two such full periods, this is speculative. Nevertheless, longer term monitoring
might yield periodicities which could be related to activity of the background blazar. Accompanied by systematic measurements
revealing the spatial fine structure of the line absorption and background continuum, this could probe the activity at the base 
of a blazar's jet (see, e.g., Nair et al. 2005), also testing the plasmon ejection model proposed by Mart\'{\i}-Vidal et al. (2013).

\begin{figure}[t]
\vspace{-0.0cm}
\centering
\resizebox{8.5cm}{!}{\includegraphics[trim=0mm -15mm 0mm 0mm]{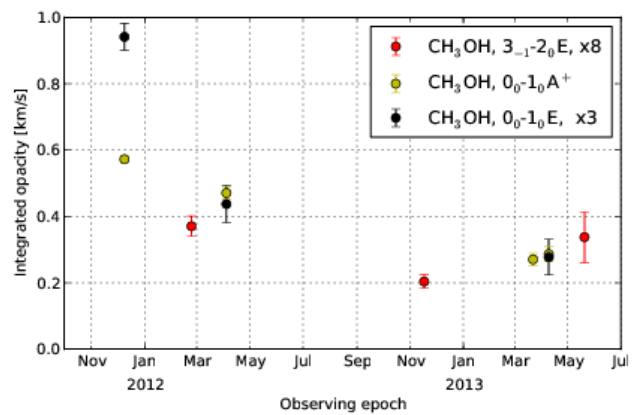}}
\vspace{-0.7cm}
\caption{Integrated opacities of three optically thin methanol (CH$_3$OH) transitions adopting a continuum source 
covering factor $f_{\rm c}$ = 0.38. The data were obtained at Effelsberg (for details, see Bagdonaite et al. 
2013a,b) and are shown to demonstrate the decline in opacities during 2012/3 in a qualitative way. Note that the 
1$_0$$\leftarrow$0$_0$ E and the 2$_0$$\leftarrow$3$_{-1}$ E opacities were multiplied by (arbitrary) factors of 3 and 8,
to bring values of all lines to roughly the same y-axis scale. Calibration uncertainties are of order 10\% or smaller.} 
\label{fig5}
\end{figure}

We thus agree with Muller et al. (2011) that the line shape experiences gradual changes which result in variations of 
velocity-integrated optical depths. Peak optical depths appear to be less variable than integrated optical depths, but we 
note that both the HCO$^+$ 2$\leftarrow$1 line (Muller et al. 2008) and, at the very center, our CS 1$\leftarrow$0 line show signs 
of saturation, while below optically thin transitions will be discussed. Nevertheless, in view of occasionally measured
broad features with questionable significance, we prefer to focus in the following on peak line intensities and the 
corresponding optical depths. The measured absorption is better defined by the depth of the (main) feature than by 
integrated opacities with sometimes poorly defined line widths. The most recent data of Muller et al. (2013) will not be 
considered because Muller et al. (2013) reported significant variations with respect to their 2009/10-spectra, in good 
agreement with our findings from CS 1$\leftarrow$0 and CH$_3$OH (see Figs.~\ref{fig3}--\ref{fig5}).

\subsection{Radiative transfer calculations}

Among our observed spectra from 2001, only SO is detected in two different transitions. However, to perform radiative transfer 
modelling for all of our observations, we need at least two different lines for each molecule. We therefore take line data from two 
other investigations from which we can derive line center optical depths, from Muller et al. (2011) for HNCO, CH$_3$OH-E, para 
and ortho c-C$_3$H$_2$ (their Table~10) and Menten et al. (1999) for ortho c-C$_3$H$_2$ and H$_2$CO. As already mentioned
in Sect.\,4.1 (see also Table~\ref{tab2}), these data are compatible, because peak opacities are similar. SO 2$_1\leftarrow1_0$ 
opacities from Muller et al. (2011) agree with our data 10 years earlier and the same is the case for c-C$_3$H$_2$ 2$_{12}\leftarrow1_{01}$ 
opacities, when comparing Menten et al. (1999) and Muller et al. (2011). 

To reproduce line ratios of the observed molecules, we apply the standard Large Velocity Gradient (LVG) model described by Henkel et 
al. (2009) assuming a spherically symmetric cloud and constant kinetic temperature and density throughout the emitting region. 
On one hand, all the molecular lines discussed below absorb less than three percent of the background continuum flux. On the other 
hand, the continuum source covering factor lies above 0.15 (Muller \& Gu{\'e}lin 2008; Henkel et al. 2009; Muller et al. 
2014b). Hence, we can be sure (as already suggested in Sect.\,4.1) that all the lines are optically thin. This implies that 
our results do not depend on cloud morphology and, therefore, our assumption of spherical cloud geometry will not affect the results. 

\begin{figure}[t]
\vspace{-0.0cm}
\hspace{-0.6cm}
%\centering
\resizebox{12.5cm}{!}{\rotatebox[origin=br]{-90}{\includegraphics[trim=0mm -15mm 0mm 0mm]{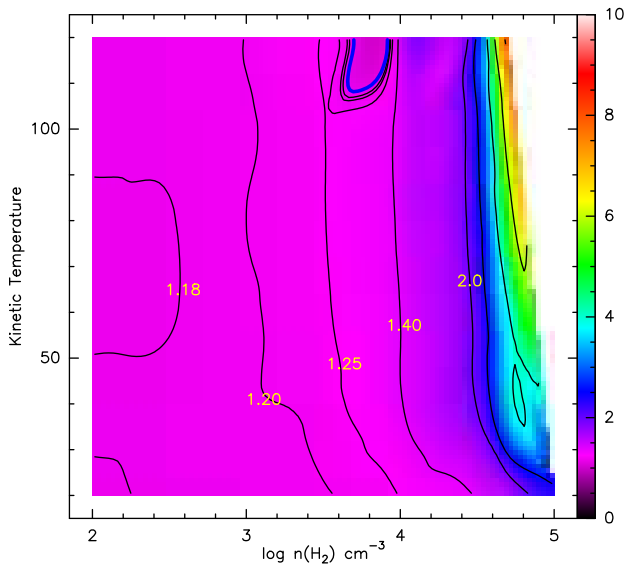}}}
\vspace{-0.7cm}
\caption{The 2$_{02}$$\leftarrow$1$_{11}$ to 1$_{11}$$\leftarrow$0$_{00}$ opacity ratio of cyclic para-cyclopropenylidene
(para c-C$_3$H$_2$) under optically thin conditions as a function of density and temperature. Depending on the continnum
source covering factors used (see subsection c-C$_3$H$_2$ in Sect.\,4.2), we obtain ratios of 0.87$\pm$0.19 and 1.14$\pm$0.25. 
Both values are below the modelled ones, but accounting for the errors, they are not inconsistent with the standard values of 
$T_{\rm kin}$ = 50--80\,K and $n$(H$_2$) = 1000--2000\,cm$^{-3}$ (Henkel et al. 2008, 2009; Muller et al. 2011, 2013).}
\label{fig6}
\end{figure}

\begin{figure}[t]
\vspace{-0.0cm}
\hspace{-0.6cm}
%\centering
\resizebox{12.5cm}{!}{\rotatebox[origin=br]{-90}{\includegraphics[trim=0mm -15mm 0mm 0mm]{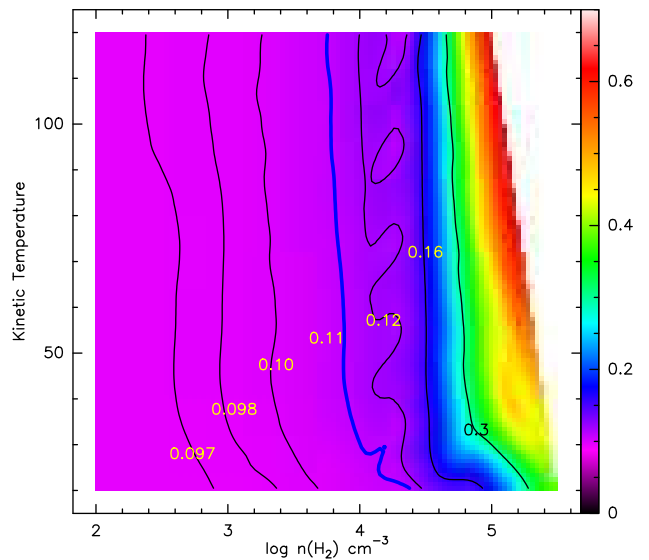}}}
\vspace{-0.7cm}
\caption{The 3$_{12}$$\leftarrow$3$_{03}$ to 2$_{12}$$\leftarrow$1$_{01}$ opacity ratio of cyclic ortho-cyclopropenylidene
(ortho c-C$_3$H$_2$) under optically thin conditions as a function of density and kinetic temperature. The blue almost 
vertical line represents the measured value (0.110$\pm$0.015; Sect.\,4.2). When accounting for the error, this is not 
inconsistent with the standard values given in the caption to Fig.~\ref{fig6}.}
\label{fig7}
\end{figure}

\begin{figure}[t]
\vspace{-0.0cm}
\hspace{-0.6cm}
%\centering
\resizebox{12.5cm}{!}{\rotatebox[origin=br]{-90}{\includegraphics[trim=0mm -15mm 0mm 0mm]{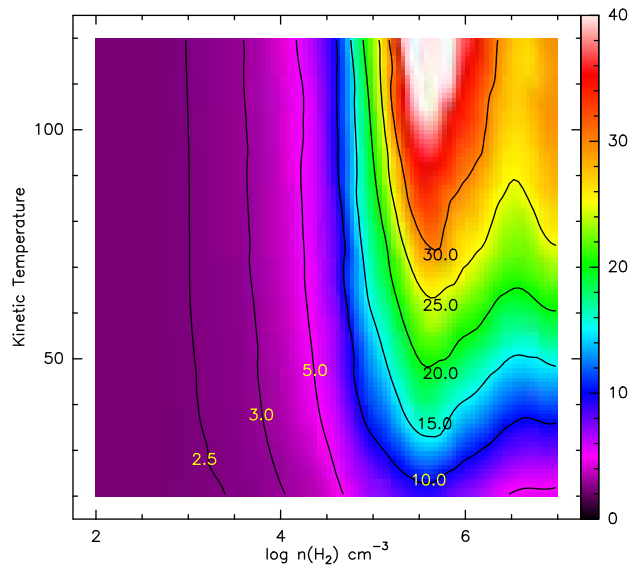}}}
\vspace{-0.7cm}
\caption{The 1$_{0}$$\leftarrow$2$_{-1}$ to 1$_{0}$$\leftarrow$0$_{0}$ opacity ratio of 
E-type methanol (CH$_3$OH) under optically thin conditions as a function of density and kinetic temperature. 
The value of 1.8$\pm$0.4 (Sect.\,4.2), deduced from measurements, indicates a low density which is, within the 
error, consistent with the standard parameters given in Fig.~\ref{fig6}.} 
\label{fig8}
\end{figure}

\begin{figure}[t]
\vspace{-0.0cm}
\hspace{-0.6cm}
%\centering
\resizebox{12.5cm}{!}{\rotatebox[origin=br]{-90}{\includegraphics[trim=0mm -15mm 0mm 0mm]{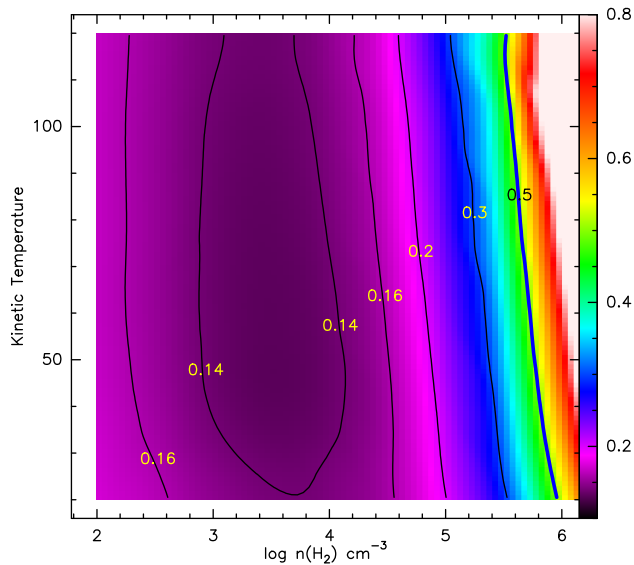}}}
\vspace{-0.7cm}
\caption{The 3$_{12}$$\leftarrow$3$_{13}$ to 2$_{11}$$\leftarrow$2$_{12}$ opacity ratio of ortho-formaldehyde 
(o-H$_2$CO) under optically thin conditions as a function of density and kinetic temperature. The blue 
thick almost vertical line on the right hand side denotes a line ratio of 0.5 (Sect.\,4.2; the ratio obtained 
with frequency dependent continuum source covering factors is 0.54$\pm$0.17). Here the density is not consistent 
with the standard value given in the caption to Fig.~\ref{fig6}.} 
\label{fig9}
\end{figure}

\begin{figure}[t]
\vspace{-0.0cm}
\hspace{-0.6cm}
%\centering
\resizebox{12.5cm}{!}{\rotatebox[origin=br]{-90}{\includegraphics[trim=0mm -15mm 0mm 0mm]{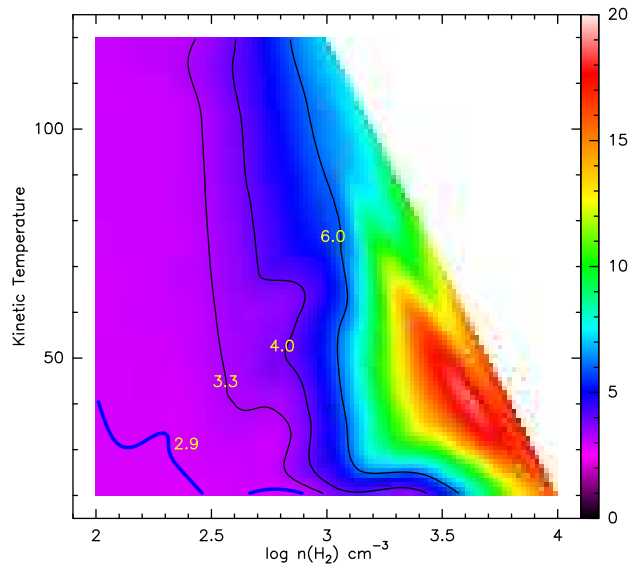}}}
\vspace{-0.7cm}
\caption{The 2$_1$$\leftarrow$1$_0$ to 1$_0$$\leftarrow$0$_1$ opacity ratio of sulfur monoxide (SO) under
optically thin conditions as a function of density and kinetic temperature. With ratios of 2.0$\pm$0.5 or
2.9$\pm$0.6, depending on the adopted continuum cource covering factors, the density is below that of the
standard value given in Fig.~\ref{fig6}. The thick blue line in the lower left corner denotes a ratio of 2.9.} 
\label{fig10}
\end{figure}

In contrast to the work of Muller et al. (2013), we do not assume single values of gas density and temperature common for all 
molecules in our analysis, which they themselves regard as the major limitation of their excitation analysis. We have carried out 
our model calculations for a large variety of densities and temperatures (see Figs.\,\ref{fig6}--\ref{fig10}) examining also 
those obtained by Henkel et al. (2008, 2009) and Muller et al. (2011).  

All data used for the model calculations are contained in Table~\ref{tab2}. The basic question is: Are the results
consistent with the density and temperature parameters of previous studies, i.e. $T_{\rm kin}$ = 50 or 80\,K and 
$n$(H$_2$) = 1000--2000\,cm$^{-3}$ (Henkel et al. 2008, 2009; Muller et al. 2011, 2013), the former involving the classical 
$T_{\rm kin}$ tracers NH$_3$ and CH$_3$CN? Because the common degeneracy between high $T_{\rm kin}$/low $n$(H$_2$) 
and vice versa affects most molecular line pairs, including our tracers, there is a wide parameter space for each pair
of lines. Here we thus intend to find out whether the above mentioned $T_{\rm kin}$/$n$(H$_2$) combination can reproduce 
our data in all cases. The results discussed below are valid for both sets of continuum coverage factors (Sect.\,4.1 and 
Table~\ref{tab2}), which slightly affect our results only in cases of significantly diverging frequencies of the lines 
involved.

\smallskip
${\bf c-C_3H_2}$ 

(p): The observed 2$_{02}\leftarrow 1_{11}$ to 1$_{11}\leftarrow 0_{00}$ ratio of para-cyclopropenylidene, 0.87$\pm$0.19
(f$_{\rm c}$ = 0.198) or 1.14$\pm$0.25 ($f_{\rm c}$ = 0.38), is -- also for the former value within 2$\sigma$ -- compatible with 
$T_{\rm kin}$ = 50--80\,K and $n$(H$_2$) = 1000--2000\,cm$^{-3}$ (Fig.\,\ref{fig6}). This agrees with the parameters derived by 
Henkel et al. (2008, 2009) and Muller et al. (2011, 2013) from multiline studies. 

(o): The measured ratio of the ortho 3$_{12}\leftarrow 3_{03}$ to 2$_{12}\leftarrow 1_{01}$ opacities, 0.110$\pm$0.015 (both lines 
taken from Muller et al. 2011), is well reproduced with the same $T_{\rm kin}$ and $n$(H$_2$) (Fig.\,\ref{fig7}) yielding a column 
density a few times that of the para-type of the molecule (the precise factor depends on the $T_{\rm kin}$ and $n$(H$_2$) parameters 
chosen), in accordance with the expected factor of three for the ortho/para ratio.   

Lines of this molecule commonly originate in moderately dense molecular gas (e.g., Thaddeus et al. 1995; Aladro et al. 
2011), and our adopted temperature and density are in accordance with this scenario. 

Although the detection of the (Galactic) C$_3$H$_2$\,(1$_{10} \leftarrow$ 1$_{01}$) feature is tentative, it should not 
be entirely unexpected since this molecular transition is known to be quite widespread (e.g., Madden et al. 1989) and our 
source is located not far from the Galactic plane and center ($l^{\rm II}$ = 12.2${^\circ}$, $b^{\rm II}$ = --5.7${^\circ}$). 
We have checked possible other line identifications in the (Galactic) range of 18.335 to 18.353\,GHz as well as in the 
range 34.581 to 34.600\,GHz, redshifted by $z$ = 0.88582, and find no other likely line candidate. There is no 
previous detection of Galactic molecular absorption lines along this line of sight. 
\smallskip

\smallskip
${\bf CH_3OH}$

Quasi-thermal lines of methanol trace warm and at least moderately dense gas ($n$(H$_2$) $\ga$ 1000\,cm$^{-3}$;
e.g., Leurini et al. 2004). The species is easily dissociated, and it is not detected in Galactic clouds of very 
low density (Liszt et al. 2008). Applying the same kinetic temperature and density as above, the modelled ratio of 
the opacities of the molecular A and E species becomes 3.8 in case of equal abundances. This agrees within 
the errors with the observations (Table~2 and Fig.~\ref{fig1}) and the more sensitive data from Bagdonaite et al. 
(2013a,b), taken more than a decade later. Referring to our clearly detected line of the A-species 
(Fig.~\ref{fig1}), dividing its intensity by 3.8, and comparing it with that of the E\,(1$_{0} \leftarrow$ 2$_{-1}$) 
transition observed by Muller et al. (2011) yields an opacity ratio of 1.8$\pm$0.4, which is also consistent 
with the density estimate by Henkel et al. (2009; see Fig.~\ref{fig8}). The kinetic temperature remains poorly 
constrained and allows for both $T_{\rm kin}$ = 50 and 80\,K. 

\smallskip
${\bf H_2CO}$

Formaldehyde can be efficiently produced on the surface of icy grains through hydrogenation of CO (e.g., Aladro et al. 2011). 
Interpolating the continuum flux density of the background source of PKS\,1830--211 at 7.7\,GHz from data at 5\,GHz and 
8.4\,GHz (Subrahmanyan et al. 1990; Nair et al. 1993; Xu et al. 1995; van Ommen et al. 1995; Mattox et al. 1997;, Lovell et 
al. 1998) yields 7.1\,Jy for the frequency of the (2$_{11} \leftarrow$ 2$_{12}$) line observed by Menten et al. (1999). Their 
absorption feature of 0.073\,Jy therefore absorbs at its peak $\sim$1\% of the total continuum flux density of PKS\,1830--211. 
From that, the derived $\tau$ becomes 0.026 for a fixed $f_{\rm c}$ = 0.38. The observed $\tau$ ratio of the 3$_{12}\leftarrow 
3_{13}$ to 2$_{11}\leftarrow 2_{12}$ lines is then of order unity while the model calculation for the temperature and 
density used above yields a ratio of 0.14 (Fig.~\ref{fig9}). To fit a ratio of 1 to 2, any temperature between 
10 and 150\,K would require densities $n$(H$_2$) $\ga$ 5$\times$10$^5$\,cm$^{-3}$. However, accounting for the fact that 
the 2$_{11} \leftarrow$ 2$_{12}$ line is measured at low frequency ($\nu_{\rm observed}$ $\sim$ 7.68\,GHz) and that therefore 
the source covering factor may be smaller, we obtain with the frequency dependent source covering factor (see Sect.\,4.1)
originally proposed by Henkel et al. (2009) $f_{\rm c}$ = 0.108 and an optical depth of 0.10. The resulting 3$_{12}\leftarrow 
3_{13}$ to 2$_{11}\leftarrow 2_{12}$ ratio of optical depths would then become $\sim$0.54$\pm$0.17, still favoring a density 
of order 10$^5$\,cm$^{-3}$ (see Fig.~\ref{fig9}), which would be consistent with the average density deduced by Mangum et al.
(2013) from the 1$_{10}$$\leftarrow$1$_{11}$ and 2$_{11}$$\leftarrow$2$_{12}$ transitions of H$_2$CO toward nearby actively star 
forming galaxies. If we adopt for our line a dip half as deep, the ratio would become $\sim$0.25--0.3 and would reduce the 
density by half an order of magnitude, which is still rather large. To bring our so far preferred model (with $T_{\rm kin}$ = 
80\,K and $n$(H$_2$) = 2000\,cm$^{-3}$) into rough accordance with the observations, our $3_{12}\leftarrow 3_{13}$ line should 
be weaker by a factor of $\sim$4, which is inconsistent with the data (see Sect.\,5 for a discussion of this result, which 
indicates strong deviations from the ``canonical'' physical parameters obtained in previous studies).

${\bf SO}$

Sulfur monoxide molecules are formed in a high temperature environment by neutral-neutral reactions with high activation 
energy (e.g., Takano et al. 1995). Both lines used for modelling (see Table~\ref{tab2}) are from our observations. The 
line shapes appear to differ from those of the other transitions displayed in Fig.~\ref{fig1} showing a redshifted line wing.
In case of the 2$_1\leftarrow 1_0$ line, a small part of the redshifted absorption might be due to the 3$_{03} \leftarrow 
2_{02}$ line of cyanic acid (HOCN), which is displaced by 28.0\,km\,s$^{-1}$ and contains hyperfine structure components
scattered over 10.7\,km\,s$^{-1}$. However, this molecule is rarely detected in the interstellar medium  (see Br{\"u}nkens 
et al. 2010) and the profile of the other SO transition (the 1$_0\rightarrow 0_1$ line) looks similar. Following Table~2,
our measured 2$_1$$\leftarrow$1$_0$/1$_0$$\leftarrow$0$_1$ peak opacity ratios are 2.0$\pm$0.5 (f$_{\rm c}$ = 0.227 
and 0.156) and 2.9$\pm$0.6 ($f_{\rm c}$ = 0.38). With $T_{\rm kin}$ = 80\,K, gas densities become very low (a few 100\,cm$^{-3}$
or less) to fit the observed line ratio (see Fig.~\ref{fig10}), which is much lower than that suggested by, e.g., Henkel et al. 
(2009). On the other hand, choosing a value near their $n$(H$_2$) = 2000\,cm$^{-3}$, the modelled gas temperature drops to 
$\la$20\,K, in strong contradiction to NH$_3$ and CH$_3$CN data (Henkel et al. 2008; Muller et al. 2011). 

The two modelled cases with different source covering factors are in accordance with the upper limit of $\tau$ determined 
from the non-detection of the 4$_3 \leftarrow$ 3$_3$ line. The only possible assumption to reconcile the result related 
to the narrow component with those from NH$_3$, HC$_3$N, and CH$_3$CN (Henkel et al. 2008, 2009; Muller et al. 2011, 2013), and our 
results from C$_3$H$_2$ and CH$_3$OH is that the absorbing molecular gas shows strong gradients at least in density and possibly 
also in $T_{\rm kin}$ (see also Sect.\,5). 
\smallskip

\begin{table*}
\label{tab4}
\caption[]{Kinetic temperatures, densities, column densities $N$, and fractional abundances $N/N_{\rm CO}$ and $N/N_{\rm H_2}$}
\begin{tabular}{lcccrrrrr}
\hline
Molecule & Temperature & Gas density & $N$ & $N_{\rm Lit.}$ & $N/N_{\rm CO}$ & $N/N_{\rm CO}$(Lit.) & $N/N_{\rm H_2}$ & $N/N_{\rm H_2}$(Lit.) \\
         &       (K)   & (cm$^{-3}$) & \multicolumn{2}{c}{(10$^{12}$\,cm$^{-2}$)} & (10$^{-6}$) & \multicolumn{1}{c}{(10$^{-6}$)} 
	            & (10$^{-9}$) & \multicolumn{1}{c}{(10$^{-9}$)}    \\
\hline
%\medskip
o+p c-C$_3$H$_2$&          80  &     2000         & 65 ($\pm$40\%) &   7  (Me) &  22 &  2  (Me) &  2.2 &  0.23  (Me)  \\
               &               &                  &                &  53  (Mu) &     & 17  (Mu) &      &  1.8   (Mu)  \\
               &               &                  &                &           &     &          &      &              \\
CH$_3$OH       &     80        &     2000         &140 ($\pm$40\%) & 170  (Mu) &  45 & 57  (Mu) &  4.5 &  5.6  (Mu)   \\
               &               &                  &                &           &     &          &      &              \\
o-H$_2$CO      &  10--150      & $>5\times10^4$   &400 ($\pm$40\%) & 380  (Me) & 130 &130  (Me) & 13.0 &  13    (Me)  \\
               &               &                  &                &  60  (Mu) &     & 20  (Mu) &      &   2    (Mu)  \\
               &               &                  &                &           &     &          &      &              \\
SO             &  20--80       & 400--2000        & 35 ($\pm$50\%) &  26  (Mu) &  12 &  9  (Mu) &  1.2 &  0.9   (Mu)  \\
               &               &                  &                &           &     &          &      &              \\
SiS            &     80        &     2000         &  $<$5.         &           &$<$2.5 &        &$<$0.2&              \\
               &               &                  &                &           &     &          &      &              \\
HNCO           &     80        &     2000         & $<$8.6         &   6.2(Mu) &$<$4 & 2.1 (Mu) &$<$0.3&  0.2  (Mu)   \\
\hline
               &               &                  &                &           &     &          &      &              \\
\end{tabular}

Col.\,(1): Molecular species;
Col.\,(2): Adopted kinetic temperature;
Col.\,(3): Adopted gas density;
Col.\,(4): Our column densities;
Col.\,(5): Column densities derived from data of Menten et al. (1999) (Me) and Muller et al. (2011) (Mu)
in the way described in Sect.\,4.3; Cols.\,(6) and (7): same as Cols.\,(4) and (5), but relative to the adopted
CO column density, $N_{\rm CO}$ = 3 $\times$ 10$^{18}$\,cm$^{-3}$ (Sect.\,4.3); Cols.\,(8) and (9): same as Cols.\,(4) 
and (5), but relative to the adopted H$_2$ column density, $N_{\rm H_2}$ = 3 $\times$ 10$^{22}$\,cm$^{-3}$ (see 
Sect.\,4.3). \\
\end{table*}

\subsection{Column densities and abundances}

Our modelled molecular column densities are presented in Table~4 as well as abundance ratios relative to CO and 
H$_2$. Our abundance ratios are generated using $N_{\rm CO}$ = 3 $\times$ 10$^{18}$\,cm$^{-2}$ (Gerin et al. 1997) 
and $N_{\rm H_2}$ = 3 $\times$ 10$^{22}$\,cm$^{-2}$ (Gerin et al. 1997, Wiklind \& Combes 1996a). This is also 
discussed by Muller et al. (2011). Depending on the background source covering factor ($f_{\rm c}$ = 0.38 or
frequency dependent; see Table~\ref{tab2}) there is in most cases an uncertainty by a factor of almost two in the 
column densities. Here we refer to average values between these two extremes, providing accordingly $\pm$40\%
errors.

To compare our results with those of Menten et al. (1999) obtained at the VLA, we take molecular column densities 
and their ratios to CO directly from their Table~1, while we derive their ratio to H$_2$ using the H$_2$ column density 
mentioned above. For comparison with results by Muller et al. (2011), we use their $N_{\rm mol}$ from their Table~3, 
and $N_{\rm CO}$  (as mentioned above) from Wiklind \& Combes (1996a) and Gerin et al. (1997). Note that our chosen 
$N_{\rm H_2}$ value is 50\% larger than that adopted by Muller et al. (2011) in their Table~8. 

${\bf c-C_3H_2}$: Our total column density is about an order of magnitude higher than that of Menten et al. (1999), 
but is in good agreement with that of Muller et al. (2011). These factors also apply to the ratios relative 
to CO and H$_2$. 

${\bf CH_3OH}$: Our column density and relative abundance are in good accordance with Muller et al. (2011) and 
also with Bagdonaite et al. (2013a,b). 

${\bf H_2CO}$: Our values agree well with those of Menten et al. (1999), while those of Muller et al. (2011)
are almost an order of magnitude smaller. 

${\bf SO}$: Our column density is in accordance with the value given by Muller et al. (2011). Their 
abundance relative to H$_2$ is almost identical to our result. For all temperatures between 20 K and 80 K, 
resulting column densities are similar within a factor of $<$2. 

${\bf SiS, HNCO}$: Our non-detections yield upper limits to the molecule's abundances relative to H$_2$. The one for HNCO
is in accordance with the value derived by Muller et al. (2011) from a detected $J$ = 3$\leftarrow$2 line.

\subsection{Excitation temperatures}

Some of the molecular lines we have measured exhibit excitation temperatures, which are quite far from the expected 
$T_{\rm CMB}$ value, $T_{\rm CMB}$ = 2.73\,K $\times$ (1+$z$) = 5.14\,K (Muller et al. 2013; Sato et al. 2013). The 
$J_{\rm KaKc}$ = 3$_{12}$$\leftarrow$3$_{13}$ line of H$_2$CO, for example, is collisionally cooled (e.g., Evans 
et al. 1975; Zeiger \& Darling 2010; Darling \& Zeiger 2012) and may show, in spite of the uncertainties w.r.t. 
density (Sect.\,4.2), excitation temperatures well below 5.14\,K. The second example is the SO 1$_0$$\leftarrow$0$_1$ 
transition. Applying our LVG code, it exhibits for the standard parameters $T_{\rm kin}$ = 80\,K and $n$(H$_2$) 
= 2000\,cm$^{-3}$ excitation temperatures of several 10\,K degrees or even inversion in case of $T_{\rm CMB}$ $<$ 4.0\,K. 
In case of lower $T_{\rm kin}$ and density, however, e.g., for 40\,K and 1000\,cm$^{-3}$ (see Sect.\,4.2), the excitation 
temperature becomes ``well behaved'' for 3\,K $<$ $T_{\rm CMB}$ $<$ 7\,K, only surpassing the $T_{\rm CMB}$ value by 
2.0--2.5\,K.  On the other hand, the LVG calculations indicate that the excitation temperatures of para and ortho 
c-C$_3$H$_2$, CH$_3$OH, and the 2$_1$$\leftarrow$1$_0$ transition of SO are normal in the sense that they are slightly 
higher than the $T_{\rm CMB}$ value and rise with increasing density (cf. Fig.~3 of Muller et al. 2013). For $n$(H$_2$) = 
2000\,cm$^{-3}$, $T_{\rm kin}$ = 80\.K, and 3\,K $<$ $T_{\rm CMB}$ $<$ 7\,K, the C$_3$H$_2$ and 1$_0$$\leftarrow$0$_0$ 
CH$_3$OH (E) excitation temperatures are within 5\% and slightly above the chosen cosmic microwave value. This emphasizes 
that these lines are suitable tools to constrain $T_{\rm CMB}$ at redshift $z$ $\sim$0.89 and confirms the assumption 
of $T_{\rm ex}$ $\ll$ $T_{\rm C}$ ($T_{\rm C}$ $\ga$ 10$^7$\,K; e.g., Jin et al. 2003), used to calculate optical depths 
(see Sect.\,3).

\section{Discussion}

The frequencies of our observations range from 15 to 35\,GHz, those of Muller et al. (2011) and Menten et al. 
(1999) from 30 to 50\,GHz with the exception of the H$_2$CO (2$_{11} \leftarrow$  2$_{12}$) line at 7.7\,GHz. 
The size of the SW spot of the background continuum source lensed by the foreground galaxy is reported to decrease 
with frequency (e.g. Carilli et al. 1998, Garrett et al. 1998, Jin et al. 2003), but the sizes given there for a 
specific frequency do not agree with each other. Nevertheless, at the highest observing frequency of 45\,GHz 
discussed in our analysis, this size must be larger than 0.2\,milliarcsec, which corresponds to $\sim$2\,pc 
linear extent.

Within the limitations by noise, all our spectra as well as the used spectra of the referenced literature 
show compatible center velocities (see, e.g., also Henkel et al. 2008 and Muller et al. 2014b). 
Therefore, we assume that they all originate from the same cloud complex. In particular, we also find that the 
line widths do not depend on the frequency of the studied feature (see also Henkel et al. 2008, 2009 
for complementary lines). Observing molecular cloud material of $\ga$2\,pc linear extent along a
line-of-sight with a likely larger depth, a line width of $\sim$10--20\,km\,s$^{-1}$  
(neglecting the $V$ $\sim$175\,km\,s$^{-1}$ component reported by Muller et al. 2014b) consisting of 
several kinematic subcomponents (e.g., Muller et al. 2008; Murphy et al. 2008; Henkel et al. 2009) 
could be readily expected by a combination of ``normal'' clouds (e.g., Larson 1981). 

From their ammonia multi-line analysis, Henkel et al. (2008) conclude that 80\% to 90\% of the ammonia 
gas has a kinetic temperature of about 80\,K. But since rotational temperatures of the ammonia gas vary from 
35\,K to up to 600\,K, they also state that the gas may show large temperature gradients. Our line analysis 
above indicates for the first time strong density gradients (for variations by a factor of two,
see also Muller et al. 2013). As Henkel et al. (2008) point out, kinetic temperatures are far too large 
for a dark-cloud-scenario. The different temperature/density parameters obtained indicate instead a 
combination of outer cloud layers and/or intercloud regions as well as denser cloud cores. 

This ``mixture'' of physical properties is also supported by a comparison of molecular abundances with 
those of different types of Galactic objects. For the following comparison of our abundance values to those 
of sources with different character, we use values of different cloud types contained in Table~8 of Muller et 
al. (2011) for our observed molecules.  There is overlap with Table~1 of Menten et al. (1999) only for the 
source TMC\,1 and for the molecules C$_3$H$_2$ and H$_2$CO; these values agree with each other. 

Our C$_3$H$_2$ abundance agrees with those of low density molecular clouds and lies between those 
of the starburst galaxies NGC\,253 and NGC\,4945. Fractional abundances for translucent clouds as well as TMC\,1 
are much higher. For CH$_3$OH, our abundance is similar to that of NGC\,4945, in between those of translucent 
clouds and TMC\,1, whereas that of NGC\,253 is a moderate factor of 2 higher. The H$_2$CO abundance is an
order of magnitude higher than in the starburst galaxies, half an order of magnitude higher than in diffuse
and translucent clouds, and is within the uncertainties (factors of two) compatible with that of TMC\,1. 
Our SO abundance agrees with those of both starburst galaxies, whereas those of translucent clouds and the 
TMC\,1 are higher. In contrast, our upper limit as well as the value for HNCO agrees with that of the TMC\,1 
and Sgr\,B2; here, abundances of translucent clouds and the starburst galaxies are higher. 

All this increasingly indicates that former interpretations attempting to attribute observed molecular 
abundances to one distinct cloud type do not lead to a realistic scenario of the interstellar matter we 
observe toward south-western molecular hotspot of PKS\,1830--211. We possibly observe a cloud 
complex which is with respect to spatial fine structure similar to cloud complexes in our own Galaxy. 
With the dust-to-gas ratio in PKS\,1830--211 being consistent with the Galactic value (Dai et al. 2006; 
Aller et al. 2012), we apparently observe a combination of different conditions in different cloud types 
which should not be surprising for a region encompassing several pc in the plane of the sky and 
presumably an even larger linear scale along the line-of-sight. 

Henkel et al. (2008), not yet being aware of the $V$ $\sim$ 175\,km\,s$^{-1}$ feature reported
by Muller et al. (2014b), discuss the scenario that the south-western absorbing cloud toward PKS\,1830--211 
may be similar to the gas seen toward Sgr\,B2 in our Galaxy. What they miss is a similar molecular 
condensation located at several kpc from the center as it is seen toward PKS\,1830--211. Line widths of various 
species observed at various linear scales in Sgr\,B2 (0.4 -- 2.5\,pc) are of order 20\,km\,s$^{-1}$ 
(Rohlfs et al. 2010, Comito et al. 2003). Apparently at $\sim$7\,Gyr in the past, also a cloud complex 
at 2--3\,kpc galactocentric distance can show this fingerprint.

\section{Conclusions}

From our excitation analysis of molecular line spectra of various species applying a Large Velocity 
Gradient radiative transfer model, we draw the following conclusions:

\begin{itemize}

\item For the c-C$_3$H$_2$ and CH$_3$OH molecules, the data are compatible with $T_{\rm kin}$ 
= 80 K and $n$(H$_2$) = 2000 cm$^{-3}$, in accordance with previously reported results, 
indicating gas of low density.

\item While our observed H$_2$CO K-doublet line is noisier than that of Menten et al. (1999), 
the resulting $\tau$ values of both lines are very similar, yielding volume densities n(H$_2$) $\ga$ 5 
$\times$ 10$^5$\,cm$^{-3}$ for all kinetic temperatures between 10 K and 80 K. An absorption dip half as 
deep as that derived from our data, which still might be adopted as an extreme, would reduce the density to n(H$_2$)  
$\sim$5 $\times$ 10$^4$\,cm$^{-3}$, which still is more than an order of magnitude higher than the one derived by
Henkel et al. (2009) on the basis of an HC$_3$N multiline analysis. 

\item For the SO molecule, if we assume T$_{\rm kin}$ = 80K, the volume gas density becomes 
extremely low (n(H$_2$) $\la$ 500 cm$^{-3}$). If we raise the density to 2000 cm$^{-3}$, the kinetic 
temperature drops to 20 K. Both models as well as those in between these two extremes result in about 
the same column density (see Table~\ref{tab4}). 

\item Our resulting densities and relative molecular abundances are in fair if not good agreement with 
those found in the local universe. This is in accordance with the statement of Aller et al. (2012) that 
``there is also evidence that the dust-to-gas ratio in this system is consistent with the Galactic value''.

\item For the south-western hotspot of PKS\,1380--211, there is a trend of rising integrated opacities 
during the late nineties, with peaks near 2001, followed by a minimum around 2006. In recent years, 
opacities were rising again, reaching a maximum in 2011/2012. In the long term we may test for 
periodicities, possibly probing a putative precessing jet in the background blazar. In this context, 
long term monitoring of both the south-western and north-eastern velocity components would be instructive.

\item Molecules are known to have different abundances in different types of clouds (e.g., Menten et al. 
1999; Muller et al. 2011). Together with the fact that the lines analyzed here reflect different physical 
conditions, we conclude that, within the observed region of a few pc in size  along the plane of the sky
and likely a larger depth along the line-of-sight, there are strong gradients in density and likely also 
in temperature. To summarize, we see a combination of conditions typical for different types of molecular gas. 

\end{itemize}

\begin{acknowledgements} We wish to thank an anonymous referee for useful comments.
 We used NASA's Astrophysical Data System (ADS), the Cologne 
 Database for Molecular Spectroscopy (CDMS; see M{\"u}ller et al. 2001, 2005), the JPL 
 Catalog (http://spec.jpl.nasa.gov/ftp/pub/catalog/catform.html), and the line lists of 
 Lovas (1992) and Coudert \& Roueff (2006). 
\end{acknowledgements}

\end{document}